\documentstyle[12pt,a41,epsfig]{article}

\def\bq{\begin{equation}}
\def\eq{\end{equation}}
\def\ba{\begin{eqnarray}}
\def\ea{\end{eqnarray}}
\def\O{{\cal{O}}}
\def\smin{s_{\mbox{\scriptsize min}}}
\def\lsimfig{\mathrel{\raise.2ex\hbox{$<$}\hskip-.8em\lower.9ex\hbox{$\sim$}}}
\def\gsimfig{\mathrel{\raise.2ex\hbox{$>$}\hskip-.8em\lower.9ex\hbox{$\sim$}}}

\def\mboxsc#1{\mbox{\scriptsize #1}}

\def\docuname{{\large \sc  mepjet}}                           
\def\docuname{{\large \sc  mepjet}}                           
  
\newcommand{\oas}{\mbox{$\mbox{$\cal{O}$}(\alpha_{s})$}}
\newcommand{\oasz}{\mbox{$\mbox{$\cal{O}$}(\alpha_{s}^{2})$}}
\newcommand{\as}{\mbox{$\alpha_{s}$}}

\begin{document}

\begin{center}
{\LARGE\bf QCD Corrections to
  Jet Production \protect\\ in Polarized {\boldmath{$ep$}} Scattering}\\
\vspace{1cm}
{E. Mirkes and S. Willfahrt}\\
\vspace*{1cm}
{\it Institut f\"ur Theoretische Teilchenphysik, 
     Universit\"at Karlsruhe,\\ D-76128 Karlsruhe, Germany}\\
\vspace*{2cm}
\end{center}
\vspace{-9cm}
\hfill \vtop{   \hbox{\bf hep-ph/9711434}
                \hbox{\bf TTP97-51}}
\vspace{9cm}
\begin{abstract}
Next-to-leading order QCD predictions for 1-jet and 2-jet 
cross sections in polarized deep inelastic scattering at HERA 
energies are presented. Whereas the QCD corrections to the total
polarized cross section are very large,
only moderate corrections are found for the dijet cross sections.
\end{abstract}

\section{Introduction}
After the confirmation of the surprising EMC result
that quarks carry  only a little fraction of the nucleon spin
the spin structure of a longitudinally proton
is  actively  being studied theoretically and
experimentally   by several
fixed target experiments
at CERN, DESY, and SLAC~\cite{ro}.
So far only the inclusive polarized structure functions
$g_1$ and $g_2$ have been measured.
These measurements, however, do not  allow to distinguish between
the role  of quarks and gluon distributions.
The measurement of the polarized gluon distribution 
$\Delta g(x_g,\mu_F ^2)$
has become the key experiment
in order to understand the 
QCD properties of the spin of the nucleon.
We study here the possible direct measurement of $\Delta g(x_g,\mu_F)$
from dijet events at a HERA collider,
in the scenario where both the electron and the proton
beam  are polarized.
The measurement of the 2-jet final state 
allows for a unique determination of the polarized
gluon distribution $\Delta g(\xi,\mu_F)$,
in a region, where $x_g \Delta g(x_g)$ is expected to show a maximum
\cite{feltesse,heraws_pol,here}.
As in the unpolarized case, the gluon distribution  enters 
the 2-jets production cross section at LO 
thus suggesting  such a direct measurement.
We will present first results of  the full NLO QCD corrections to
the 2-jet cross section in this contribution.
The numerical results are based on the
fully differential $ep \rightarrow n$ jets event generator \docuname\ 2.2 
\cite{plb1,habil}
which allows to analyze any 1- or 2-jet  like
observable in polarized $ep-$scattering in NLO 
in terms of parton 4-momenta.

The NLO hadronic $n$-jet cross section is given by
\begin{equation}
d\Delta\sigma_{\mboxsc{had}}[n\mbox{-jet}] = 
\sum_a \int d x_a \,\,\Delta f_a(x_a,\mu_F)\,\,\,
\as^n(\mu_R) \,\, \Delta \hat{\sigma}_a(p_0=x_a P, \mu_R, \mu_F)
\label{sighaddefpol}
\end{equation}
Here, the polarized hadronic cross section is defined as
\begin{equation}
d\Delta\sigma_{\mboxsc{had}}[n\mbox{-jet}] \equiv
\frac{1}{2}
\left(
d\sigma_{\mboxsc{had}}^{\uparrow\downarrow}[n\mbox{-jet}] 
-
d\sigma_{\mboxsc{had}}^{\uparrow\uparrow}[n\mbox{-jet}] 
\right)
\end{equation}
where the left arrow in the superscript denotes the polarization of the 
incoming lepton with respect to the direction of its momentum.
The right arrow stands for the polarization of the proton parallel 
or anti-parallel to the polarization of the incoming lepton.
The polarized parton distributions are defined by
$\Delta f_a(x_a,\mu_F^2)
\equiv f_{a \uparrow}(x_a,\mu_F)-f_{a \downarrow}(x_a,\mu_F)$.
Here, $f_{a \uparrow} (f_{a \downarrow})$ denotes 
the probability to find a parton $a$ 
in the longitudinally polarized  
proton whose spin is aligned (anti-aligned) to the proton's spin.

$\Delta\hat{\sigma}_a$ 
denotes the polarized  NLO differential  partonic cross section 
for the subprocess
\begin{equation}
e^\pm(l) + \mbox{parton a}(p_0) \rightarrow  e^\pm(l^\prime)+   
\mbox{parton} \,\,1 (p_1)
\ldots
+\mbox{parton}\,\, n (p_n)
\end{equation}
with $\as$ set to one from which collinear initial state 
singularities have been factorized out 
at a scale $\mu_F$ and have been
implicitly included in the scale dependent parton densities 
$f_a(x_a,\mu_F)$.
The following tree level and one loop subprocesses contribute to 
polarized $n$-jet production (up to NLO for $n=1,2$)
\begin{equation}
\begin{array}{lllll}
\mbox{1-jet:} 
& \mbox{LO} 
& \O(\alpha_s^0): 
& e+q\rightarrow e + q 
&  \mbox{}   \\[1mm]
\mbox{1-jet:}
& \mbox{NLO} 
& \O(\alpha_s):
& e+q\rightarrow e + q 
& \hspace{-4cm} \mbox{1-loop corrections}  \\
 \mbox{}
& 
&
& \mbox{+ unresolved contributions from the $\O(\as)$}
& \hspace{-2mm}\mbox{2-parton final states }  \\
\mbox{2-jets:} 
& \mbox{LO} 
& \O(\alpha_s): 
& e+q\rightarrow e + q + g 
&   \\
  \mbox{}
&
& 
& e+g\rightarrow e + q + \bar{q} 
&   \\
  \mbox{2-jets:}
& \mbox{NLO} 
& \O(\alpha_s^2):
& e+q\rightarrow e + q + g
&\hspace{-4cm} \mbox{1-loop corrections}  \\
\mbox{}
&
& 
& e+g\rightarrow e + q + \bar{q} 
&\hspace{-4cm} \mbox{1-loop corrections}  \\
  \mbox{}
&
& 
& \mbox{+ unresolved contributions from the $\O(\alpha_s^2)$}
& \hspace{-2mm}\mbox{3-parton final states }  \\
  \mbox{3-jets:} 
& \mbox{LO} 
& \O(\alpha_s^2): 
& e+q\rightarrow e + q + g + g 
&    \\
  \mbox{}
&
& 
& e+q\rightarrow e + q + q + \bar{q}
&   \\
  \mbox{}
&
& 
& e+g\rightarrow e + q + \bar{q} + g
    \\
\end{array}
\label{processes}
\end{equation}
and the crossing related anti-quark processes with $q\leftrightarrow \bar{q}$.

First discussions about jet
production in polarized lepton-hadron scattering
can be found in Ref.~\cite{ziegler}, where jets were defined in
the JADE scheme 
for center of mass energies of 20~GeV,
(which is about the energy of the fixed-target  EMC  experiment
at CERN  with a polarized muon beam of enery around 220 GeV).
Although this energy is  too small to observe clear jet structures, 
the studies in Ref.~\cite{ziegler} demonstrated alredy the unique possibility
for a measurement of the polarized gluon density from dijet events in
polarized DIS.
Prospects for measuring the polarized gluon distribution at HERA energies 
have been discussed first in  \cite{feltesse,heraws_pol}.
The results have been confirmed with the PEPSI program \cite{pepsi}.
An extension of these studies 
including a first discussion of the QCD corrections to the polarized
inclusive and 1-jet cross sections and \oasz\ polarized 3-jet cross sections
was presented in Ref~\cite{habil}.
In this contribution, we present first NLO results for
dijet cross sections.

\subsection{Polarized Jet Cross Sections\protect\vspace{1mm}}
Following the theoretical framework for the  calculation
of NLO jet cross section in unpolarized $ep-$scattering
as explained in depth in Ref.~\cite{habil},
the polarized hadronic 1-jet {\it exclusive}
cross section in the 1-photon exchange up to $\O(\alpha_s)$ reads
\\[5mm]
\fbox{\rule[-5cm]{0cm}{8cm}\mbox{\hspace{16.2cm}}}
\\[-8.3cm]
\begin{eqnarray}
\displaystyle
\Delta\sigma_{\mboxsc{had}}[\mbox{1-jet}] &=&
\nonumber\\
&&\hspace{-15mm}
\int_0^1d\eta \int
\,d{\mbox{PS}}^{(l^\prime+1)}\,\,\sigma_0\,\,
\bigg[  \nonumber \\
&&  [\sum_{i=q,\bar{q}}e_i^2 \Delta f_i(\eta,\mu_F)\big]\,
\,\,\Delta |M^{(\mboxsc{pc})}_{q\rightarrow q}|^2\,
\left(1+\alpha_s(\mu_R)\,
{\cal{K}}_{q\rightarrow q}(\smin,Q^2)\right)\nonumber \\
&+&
\,
[\sum_{i=q,\bar{q}}e_i^2\, 
\,\Delta C_i^{\overline{\mboxsc{MS}}}(\eta,\mu_F,\smin)]\,\,
\alpha_s(\mu_R)\,\,\,\Delta |M^{(\mboxsc{pc})}_{q\rightarrow q}|^2\,
\,\bigg] J_{1\leftarrow 1}(\{p_i\}) 
\label{onejetpol}\\
&&\hspace{-20mm}
+
\int_0^1 d\eta \int
\,d\mbox{PS}^{(l^\prime+2)}\,\,\sigma_0\,\,
(4\pi\alpha_s(\mu_R))\,\,  
\bigg[    \nonumber \\
&&[\sum_{i=q,\bar{q}}e_i^2 \Delta f_i(\eta,\mu_F)] \,
\,\,\Delta |M^{(\mboxsc{pc})}_{q\rightarrow qg}|^2
\nonumber    \\ 
&+&
(\sum_{i=q}e_i^2 ) \Delta f_g(\eta,\mu_F)\,\,
\,\Delta |M^{(\mboxsc{pc})}_{g\rightarrow q\bar{q}}|^2
\bigg]
\,\,
\prod_{i<j;\,0}^{2}\Theta(|s_{ij}| - \smin)\,\,
J_{1\leftarrow 2}(\{p_i\})
\nonumber\\[5mm]\nonumber
\end{eqnarray}
where the Lorentz-invariant phase space measure
$d\mbox{PS}^{(l^\prime+n)}$
is defined in Eq.~(9) of Ref.~\cite{habil}
and $\sigma_0 = \pi^2\alpha^2/((p_0.l) Q^4$).
$J_{1\leftarrow 1}(\{p_i\})$ and $J_{1\leftarrow 2}(\{p_i\})$ 
represents the jet algorithms.
The jet algorithm $J_{n \leftarrow n}$ yields one if the
original final state $n$-parton configuration yields $n$ jets satisfying the
experimental cuts. $J_{n \leftarrow n}$ 
can be expressed as a product of a clustering
part and an acceptance part \cite{habil}.
Similarly, the jet algorithm 
$J_{n\leftarrow n+1}$ evaluates to one if the $(n+1)$-
parton configuration yields $n$ detected jets and vanishs otherwise.
More precisely, $J_{n\leftarrow n+1}$ 
evaluates to one either if one pair
of partons is clustered into one jet and the remaining $(n-1)$ partons
are well separated from this jet and pass all acceptance criteria 
(together with the jet) or all $(n+1)$ partons are resolved but one parton 
does not pass the acceptance cut.
For more details see Eqs.~(10)
and~(18), respectively in Ref.~\cite{habil}.
The 1-jet {\it inclusive} cross section is defined via
Eq.~(\ref{onejetpol}) by replacing
$J_{1\leftarrow 2}(\{p_i\})$ in the last line by 
($J_{1\leftarrow 2}(\{p_i\})+J_{2\leftarrow 2}(\{p_i\}))$,
{\it i.e.} the 1-jet inclusive cross section is defined
as the sum of the NLO 1-jet exclusive cross section (as defined
in Eq.~(\ref{onejetpol})) plus the LO two jet cross section.
The theoretical cutoff parameter $\smin$ is a completely
unphysical parameter and the numerical results for any infrared safe
observable are insensitive to a reasonable variation for
sufficiently small $\smin$ values \cite{plb1,habil}.
The dynamical ${\cal{K}}_{q\rightarrow q}$  factor
in Eq.~(\ref{onejetpol}) is the same as is 
unpolarized case given in Eq.~(164) in Ref~\cite{habil}.
The polarized squared matrix elements   in Eq.~(\ref{onejetpol}) are 
\cite{habil}:
\begin{eqnarray}
\Delta |M^{(\mboxsc{pc})}_{q\rightarrow q}|^2&=&
|M^{(\mboxsc{pv})}_{q\rightarrow q}|^2=
32\,
\left[
(p_0.l)^2  \, -\,
(p_0.l^\prime)^2 \,
\right]
\,= \,8 \hat{s}^2\,\,(1-(1-y)^2)
\label{dm_qtoq} \\[2mm]
\Delta |M^{(\mboxsc{pc})}_{q\rightarrow qg}|^2
&=&
|M^{(\mboxsc{pv})}_{q\rightarrow qg}|^2 
=
\frac{128}{3}\,\,(l.l^\prime)\,\,
\frac{(l.p_0)^2-(l^\prime.p_0)^2-(l.p_1)^2+(l^\prime.p_1)^2
     }{(p_1.p_2)(p_0.p_2)}
\label{dm_qtoqg} \\[2mm]
\Delta |M^{(\mboxsc{pc})}_{g\rightarrow q\bar{q}}|^2
&=&
16\,\,(l.l^\prime)\,
\frac{(l^\prime.p_2)^2-(l.p_2)^2+(l^\prime.p_1)^2-(l.p_1)^2
     }{(p_0.p_1)(p_0.p_2)}
\label{dm_gtoqqbar} 
\end{eqnarray}
Color facors (including the initial state color average) 
are included in these results.

The structure of the polarized crossing functions
$\Delta C_{q,\bar{q}}^{\overline{\mboxsc{MS}}}(\eta,\mu_F,\smin)$
in Eq.~(\ref{onejetpol})
is  identical to the structure of the unpolarized crossing functions
\cite{giele2,habil}:
\begin{equation}
\Delta C_{a}^{\overline{\mboxsc{MS}}}(x,\mu_F,\smin)=
\left(\frac{N}{2\pi}\right)
\left[ \Delta A_{a}(x,\mu_F)\ln\left(\smin/\mu_F\right)
+      \Delta B_{a}^{\overline{\mboxsc{MS}}}(x,\mu_F)\right]
\label{dcrossf}
\end{equation}
with
\begin{equation}
\Delta A_a(x,\mu_F) = \sum_p \Delta A_{p\rightarrow a}(x,\mu_F)
%
%
\hspace{1.5cm}
\Delta B_a^{{\overline{\mboxsc{MS}}}}(x,\mu_F) = \sum_p
\Delta B_{p\rightarrow a}^{{\overline{\mboxsc{MS}}}}(x,\mu_F)
\end{equation}
The sum runs  over $p=q,\bar{q},g$.
More specifically, the polarized  crossing functions for valence quarks
and sea quarks, which are needed in Eq.~(\ref{onejetpol}),
can be obtained from Eqs.~(27,28) in Ref.~\cite{habil} by replacing
$A,B,C$ by $\Delta A,\Delta B,\Delta C$ respectively.
The finite  functions 
$\Delta A_{q\rightarrow q}(x,\mu_F)$ 
and 
$\Delta B_{q\rightarrow q}^{{\overline{\mboxsc{MS}}}}(x,\mu_F)$ 
can be obtained from  the r.h.s. of
Eqs.~(31,35) in Ref.~\cite{habil} with $f_q$ replaced by $\Delta f_q$.
The polarized $g\rightarrow q$ induced functions
$\Delta A_{g\rightarrow q}(x,\mu_F)$ 
and
$\Delta B_{g\rightarrow q}^{{\overline{\mboxsc{MS}}}}(x,\mu_F)$ 
are given by
\begin{eqnarray}
\Delta A_{g\rightarrow q} &=& \int_x^1 \frac{dz}{z}\,\, \Delta f_g(x/z,\mu_F)
\,\,\,\frac{1}{4}\,\,\,\Delta\hat{P}^{(4)}_{g\rightarrow q}(z)
\label{dagq}
\\
\Delta B_{g\rightarrow q}^{{\overline{\mboxsc{MS}}}} &=& 
\int_x^1 \frac{dz}{z}\,\,
\Delta f_g(x/z,\mu_F)
\,\,\,\frac{1}{4}\,\,\,\left\{ \Delta\hat{P}^{(4)}_{g\rightarrow q}(z)
\ln(1-z)-\Delta \hat{P}^{(\epsilon)}_{g\rightarrow q}(z)\right\}
\label{dbgq}
\end{eqnarray}
with the polarized Altarelli-Parisi kernels 
\begin{eqnarray}
\Delta\hat{P}^{(4)}_{g\rightarrow q}(z) &=& 
\frac{1}{3} 
\Delta P^{(4)}_{q\bar{q}\rightarrow g}(z) = 
\frac{2}{3}\,[2z-1] \\
\Delta \hat{P}^{(\epsilon)}_{g\rightarrow q}(z) &=& 
\frac{1}{3} 
\Delta P^{(\epsilon)}_{g\rightarrow q}(z) = 
0
\end{eqnarray}
The corresponding formula to Eq.~(\ref{onejetpol}) for the
NLO 2-jet cross section is too long to be presented here.

\subsection{Numerical Results\protect\vspace{1mm}}
In this part we present numerical results for polarized NLO
1- and 2-jet cross sections at HERA energies.
Eq.~(\ref{onejetpol}) includes all relevant information
for the calculation of the fully differential 1-jet or total \oas\
polarized cross section.

For the following numerical studies
we use a cone algorithm 
defined in the lab frame with $\Delta R=1$
and $p_T^{\mboxsc{lab}}>5$ GeV.
Events are selected
in the $Q^2$ range of  $40<Q^2< 2500$ GeV$^2$.
In addition, we require 
$0.3 < y < 1$,
an energy cut of $E(e^\prime)>5$~GeV on the scattered 
electron, and a cut on the pseudo-rapidity $\eta=-\ln\tan(\theta/2)$
of the scattered lepton and jets of $|\eta|<3.5$. 
The renormalization scale  and
the factorization scale  are set to $Q$.
The results are based on 
parton distributions from Gehrman and Stirling (GS)
\cite{gs} ``gluon set A'' together with
the 2-loop formula for the strong coupling constant.

With these parameters, one obtains 
23~pb for the LO 1-jet cross section 
$(\equiv \sigma_{\mboxsc{tot}}^{\mboxsc{LO}})$
and -10.4~pb for the NLO
1-jet inclusive cross section\footnote{The difference between
the 1-jet (inclusive)
cross section and the total $({\cal{O}}(\alpha_s))$ cross section
due to effects discussed in section~4.2 in Ref.~\protect\cite{habil}
are small and will be neglected in the following.}
$(\equiv \sigma_{\mboxsc{tot}}^{\mboxsc{NLO}})$.

\begin{figure}[htb]
  \centering
  \mbox{\epsfig{file=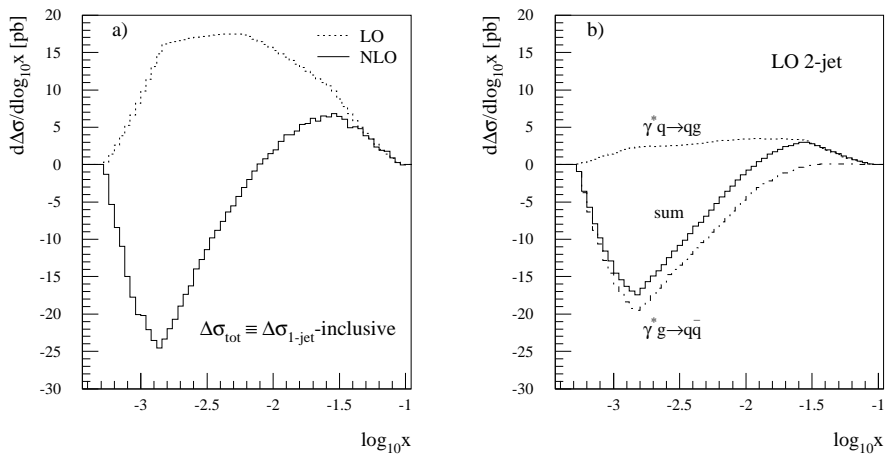,bbllx=0,bblly=270,bburx=550,bbury=550,
         width=0.8\linewidth}} 
\vspace*{-3mm}
\caption{
a) Dependence of the polarized LO and NLO 1-jet inclusive
cross section as a function of Bjorken $x$  with cuts as decribed
in the text. LO (NLO) results are based on
LO (NLO) ``gluon set A'' parton distributions  \protect\cite{gs};
b) LO 2-jet contribution
to the NLO 1-jet inclusive result in a). 
Jets are required to have  $p_T^{\protect\mboxsc{lab}}(j)>5$ GeV.
Results are shown for the quark and gluon initiated subprocesses 
alone and for the sum.
}
\label{f_pol1}
\end{figure}

The origin of these extremely  large corrections is investigated in 
Fig.~\ref{f_pol1}. The Bjorken-$x$ dependence of the corresponding
cross sections  in Fig.~\ref{f_pol1}a shows that the corrections are 
dominated by events at small $x$.
As already mentioned the  \oas\ corrected 1-jet inclusive 
cross section (solid curve in Fig.~\ref{f_pol1}a) 
is defined as the sum of the NLO 1-jet exclusive
and the LO 2-jet cross section.
Fig.~1b shows the $x$ dependence of the hard LO 2-jet
contribution.
The negative corrections are entirely
due to the hard boson-gluon fusion subprocess (lower curve
in Fig.~\ref{f_pol1}b), which is negative for 
$x\lsimfig 0.025$, whereas the contribution from the
quark-initiated process is positive (but fairly small) 
over the whole kinematical range.
The important observation is that
the \oas\ corrections in Fig.~\ref{f_pol1}a
are dominated by these \underline{hard} 2-jet events,
and in particular by the large negative 
contribution from the boson-gluon fusion subprocess.
 
\begin{figure}[htb]
  \centering
  \mbox{\epsfig{file=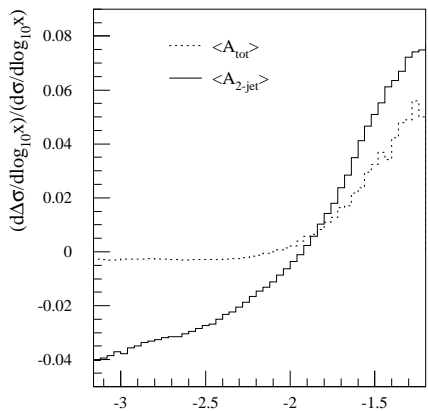,bbllx=0,bblly=400,bburx=300,bbury=650,
         width=0.45\linewidth}} 
\vspace*{-10mm}
\caption{
Asymmetries $\langle A_{\mboxsc{2-jet}} \rangle$
and $\langle A_{\mboxsc{tot}} \rangle$ in Eq.~(\protect\ref{asymdef})
as a function of $x$.
}
\label{f_pol2}
\end{figure}

In order to compare the feasibility and the
sensitivity of the  measurement of the spin asymmetry at HERA
energies, Fig.~\ref{f_pol2}a compares the 
asymmetries 
\begin{equation}
\langle A_{\mboxsc{tot}} \rangle = 
\frac{\Delta\sigma^{\mboxsc{had}}_{\mboxsc{NLO}}[\mbox{tot}]}{
            \sigma^{\mboxsc{had}}_{\mboxsc{NLO}}[\mbox{tot}]}
\hspace{2cm}
\langle A_{\mboxsc{2-jet}} \rangle = 
\frac{\Delta\sigma^{\mboxsc{had}}[\mbox{2-jet}]}{
            \sigma^{\mboxsc{had}}[\mbox{2-jet}]}
\label{asymdef}
\end{equation}
as a function of $x$.
The unpolarized cross sections in the denumerators of Eq.~(\ref{asymdef})
are based on NLO GRV \cite{grv} parton distribution functions.
\begin{figure}[htb]
\vspace*{1cm}
  \centering
  \mbox{\epsfig{file=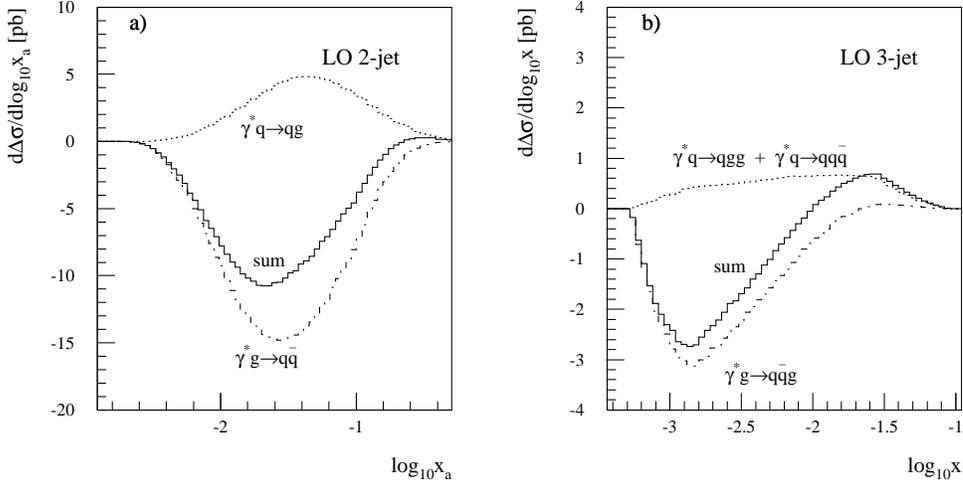,bbllx=40,bblly=400,bburx=540,bbury=630,
         width=0.8\linewidth}} 
\vspace*{-5mm}
\caption{
a)
Same as Fig.~\protect\ref{f_pol1}b for the $x_a$ ($a=q,g$) distribution, 
$x_a$ representing
the momentum fraction of the incident parton at LO;
b) \protect\oasz\ 3-jet 
contribution with $p_T^{\protect\mboxsc{lab}}(j)>5$ GeV.
Results are shown for the quark and gluon initiated subprocesses 
alone and for the sum.
}
\label{f_pol3}
\end{figure}
\begin{figure}[thb]
\vspace*{1cm}
  \centering
  \mbox{\epsfig{file=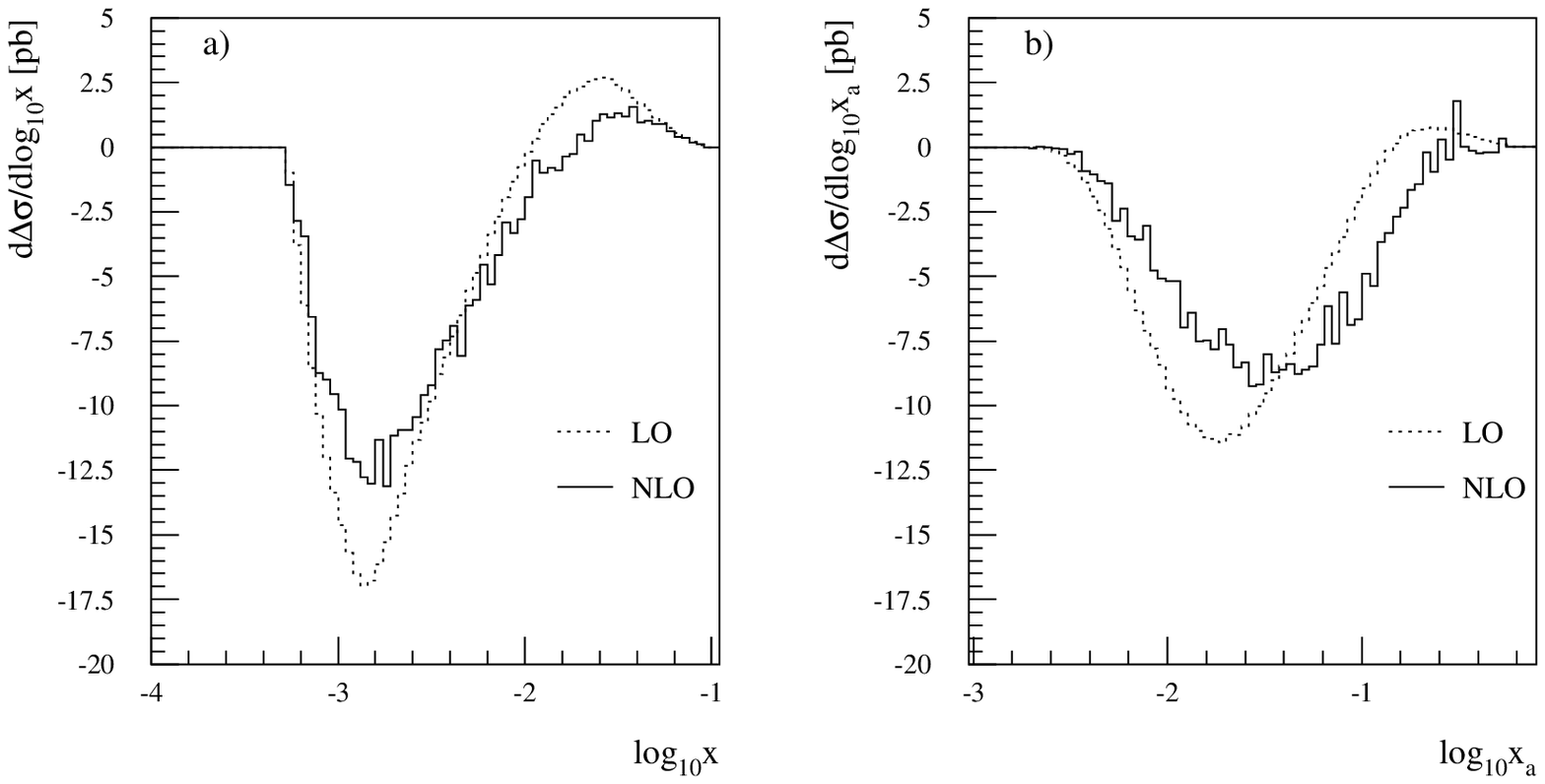,bbllx=0,bblly=270,bburx=540,bbury=540,
         width=0.8\linewidth}} 
\vspace*{-5mm}
\caption{
(a) Dependence of the polarized exclusive 2-jet cross section
 on Bjorken $x$.
Both LO (dashed) and NLO (solid) results are shown.
LO (NLO) results are based on
LO (NLO) ``gluon set A'' parton distributions  \protect\cite{gs}.
(b) Same as a) for the $x_i$ distribution, $x_i$ representing the 
momentum fraction of the incident parton at LO.
}
\label{f_pol4}
\end{figure}

One observes that the dijet asymmetry $\langle A_{\mboxsc{2-jet}} \rangle$
is much larger (up to 3-4\%) than the inclusive asymmetry
$\langle A_{\mboxsc{tot}} \rangle$ in the
low $x$ region, which is hardly (or even not at all)
constrained by currently available DIS data.
Thus, the dijet events from polarized electron and
polarized proton collisions at HERA 
are expected to provide the best measurement of the gluon polarization
distribution in the small $x$ regime.

For the isolation of polarized 
parton structure functions we are interested, however,  in the 
fractional momentum $x_a$ of incoming parton $a$ ($a=q,g$),
which is related to $x$ by 
\begin{equation}
x_a = x \,\left(1+{{s_{jj}}}/{Q^2}\right)
\label{xadef}
\end{equation}
($s_{jj}$ denotes the 
invariant mass squared of the two jets).
The corresponding $x_a$ distributions of the polarized 2-jet cross 
sections are shown in Fig.~\ref{f_pol3}a.

LO predictions for the polarized \oasz\  3-jet cross sections are shown
as a function of Bjorken-$x$ in Fig.~\ref{f_pol3}b.
One observes a very similar shape for the gluon
and quark initiated subprocesses as already found 
for the 2-jet results in Fig.~\ref{f_pol1}b.
Moreover, the 3-jet cross sections are now suppressed
by about a factor five compared to the LO 2-jet cross sections
in Fig.~\ref{f_pol1}b.
Note that the new gluon initiated subprocess $eg\rightarrow eq\bar{q}$ 
was responsible for the very large \oas\ corrections
in the 1-jet inclusive case. 
There is no such {\it new } contributing
subprocess starting at \oasz\ (see Fig.~\ref{f_pol3}b), which could
introduce  similarly large corrections to the 2-jet results.

\begin{figure}[htb]
  \centering
  \mbox{\epsfig{file=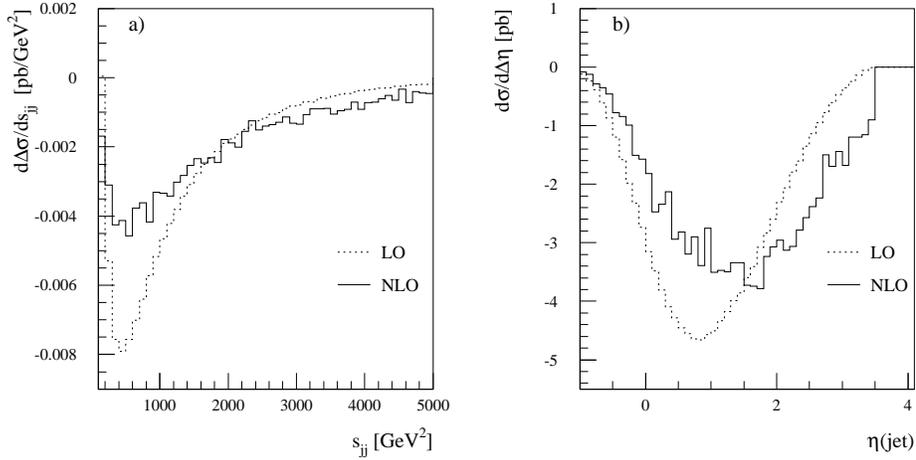,bbllx=0,bblly=270,bburx=540,bbury=540,
         width=0.8\linewidth}} 
\vspace*{-5mm}
\caption{
(a) Dijet invariant mass distribution in LO (dashed) and in NLO (solid).
(b) Rapidity distribution of the most forward jet 
    in the lab frame. Results are shown in LO
    (dashed curves) and NLO (solid) for the 
    2-jet exclusive cross section.
}
\label{f_pol5}
\end{figure}

The QCD corrections to the dijet cross sections are investigated in
Figs.~\ref{f_pol4},\ref{f_pol5}.
Compared to the previous results for dijet cross sections,
jets are required to have transverse momenta of at least 5 GeV in the
laboratory frame \underline{and} in the Breit frame.
Furthermore the renormalization scale  and
the factorization scale  are set to $\mu_R=\mu_F= 1/2\,\sum_j \,k_T^B(j)$
\cite{habil}.
Here $(k_T^{B}(j))^2$ is defined by 
$2\,E_j^2(1-\cos\theta_{jP})$, where the subscripts $j$ and $P$
denote the jet and proton, respectively (all quantities are defined
in the Breit frame).
LO (NLO) results are based on LO (NLO) parton distributions from Ref.~\cite{gs}
with the 2-loop formula for the strong coupling constant.
With these parameters 
one obtains a LO (NLO) polarized
two jet cross section $\Delta\sigma^{\mboxsc{had}}(2\mbox{-jet})$
of -10.4 pb (-10.0 pb).
Thus the higher order corrections are small. 

Fig.~\ref{f_pol4}a  
shows the dependence of the polarized exclusive 2-jet cross section
in the cone scheme on Bjorken $x$ in LO and NLO.
The effective $K$-factor close to unity
for the  exclusive dijet cross section is a consequence of
compensating effects in the low $x$ ($K<$ 1) and high $x$ ($K>$1) regime.
The dependence on the fractional 
momentum $x_a$ of incoming parton $a$ ($a=q,g$),
which is related by Eq.~(\ref{xadef}) to Bjorken$-x$,
is shown in Fig.~\ref{f_pol4}b.
The NLO distribution is shifted towards larger $x_a$ values
compared to the LO result.
The  $s_{jj}$ distribution in Fig.~\ref{f_pol5}a
exhibits fairly large QCD corrections
as well. In particular, the invariant mass squared of the two jets is
larger at NLO than at LO.

Another observable which exhibits rather large NLO corrections is the 
the jet rapidity of the most forward jet in the 
lab frame shown in Fig.~\ref{f_pol5}b.
At NLO jets are produced somewhat more 
forward (in the proton direction) than at LO.
Hence, the rapidity cut at $|\eta_j|=3.5$ has a stronger effect in NLO,
which partially explains the 
$K$-factor close to one.

We found, that the QCD corrections to the transverse momentum
distributions of the jets in both the lab and Breit frame
are fairly small.
Furthermore, it is shown in Refs.~\cite{feltesse,heraws_pol,here} that
the 2-jet spin asymmetry is not washed out by hadronization
effects. 
Asymmetry distributions for additional kinematical variables 
are considered in Ref.~\cite{maul}.

In  conclusion,  the QCD corrections to 
dijet cross sections in polarized  electron
and polarized proton collisions at HERA are found to be  moderate.
Thus, dijet events can provide
a good measurement of the gluon polarization
distribution.

\def\ap#1#2#3   {{\em Ann. Phys. (NY)} {\bf#1} (#2) #3}
\def\apj#1#2#3  {{\em Astrophys. J.} {\bf#1} (#2) #3}
\def\apjl#1#2#3 {{\em Astrophys. J. Lett.} {\bf#1} (#2) #3}
\def\app#1#2#3  {{\em Acta. Phys. Pol.} {\bf#1} (#2) #3}
\def\ar#1#2#3   {{\em Ann. Rev. Nucl. Part. Sci.} {\bf#1} (#2) #3}
\def\cpc#1#2#3  {{\em Computer Phys. Comm.} {\bf#1} (#2) #3}
\def\err#1#2#3  {{\it Erratum} {\bf#1} (#2) #3}
\def\ib#1#2#3   {{\it ibid.} {\bf#1} (#2) #3}
\def\jmp#1#2#3  {{\em J. Math. Phys.} {\bf#1} (#2) #3}
\def\ijmp#1#2#3 {{\em Int. J. Mod. Phys.} {\bf#1} (#2) #3}
\def\jetp#1#2#3 {{\em JETP Lett.} {\bf#1} (#2) #3}
\def\jpg#1#2#3  {{\em J. Phys. G.} {\bf#1} (#2) #3}
\def\mpl#1#2#3  {{\em Mod. Phys. Lett.} {\bf#1} (#2) #3}
\def\nat#1#2#3  {{\em Nature (London)} {\bf#1} (#2) #3}
\def\nc#1#2#3   {{\em Nuovo Cim.} {\bf#1} (#2) #3}
\def\nim#1#2#3  {{\em Nucl. Instr. Meth.} {\bf#1} (#2) #3}
\def\np#1#2#3   {{\em Nucl. Phys.} {\bf#1} (#2) #3}
\def\pcps#1#2#3 {{\em Proc. Cam. Phil. Soc.} {\bf#1} (#2) #3}
\def\pl#1#2#3   {{\em Phys. Lett.} {\bf#1} (#2) #3}
\def\prep#1#2#3 {{\em Phys. Rep.} {\bf#1} (#2) #3}
\def\prev#1#2#3 {{\em Phys. Rev.} {\bf#1} (#2) #3}
\def\prl#1#2#3  {{\em Phys. Rev. Lett.} {\bf#1} (#2) #3}
\def\prs#1#2#3  {{\em Proc. Roy. Soc.} {\bf#1} (#2) #3}
\def\ptp#1#2#3  {{\em Prog. Th. Phys.} {\bf#1} (#2) #3}
\def\ps#1#2#3   {{\em Physica Scripta} {\bf#1} (#2) #3}
\def\rmp#1#2#3  {{\em Rev. Mod. Phys.} {\bf#1} (#2) #3}
\def\rpp#1#2#3  {{\em Rep. Prog. Phys.} {\bf#1} (#2) #3}
\def\sjnp#1#2#3 {{\em Sov. J. Nucl. Phys.} {\bf#1} (#2) #3}
\def\spj#1#2#3  {{\em Sov. Phys. JEPT} {\bf#1} (#2) #3}
\def\spu#1#2#3  {{\em Sov. Phys.-Usp.} {\bf#1} (#2) #3}
\def\zp#1#2#3   {{\em Zeit. Phys.} {\bf#1} (#2) #3}

\end{document}